\documentclass[a4paper,10pt,twoside]{cpc-hepnp}

\usepackage{multicol}
\usepackage{graphicx}
\usepackage{booktabs}
\usepackage{amssymb,bm,mathrsfs,bbm,amscd}
\usepackage[tbtags]{amsmath}
\usepackage{lastpage}
\usepackage{upgreek}
\usepackage{CJK}
\usepackage{enumerate}
\usepackage{flushend,cuted}

\begin{document}
\begin{CJK*}{GBK}{song}

\fancyhead[c]{\small Chinese Physics C~~~Vol. XX, No. X (201X)
XXXXXX} \fancyfoot[C]{\small 010201-\thepage}

%\footnotetext[0]{Received 14 March 2009}
\footnotetext[0]{* The study was supported by the Joint Funds of the NSFC under Contracts No. U1232105 and funding from CAS for the Hundred Talent programs No. Y3515540U1.}

\title{Cross Section and Higgs Mass Measurement with Higgsstrahlung at the CEPC}

\author{%
      %Author1$^{1,2;1)}$\email{wuyf@mail.ihep.ac.cn}%
\quad Zhen-Xing Chen $^{1,2;1}$\email{zxchen@ihep.ac.cn}%
\quad Ying Yang $^{2}$%\email{yangying@ihep.ac.cn}%
\quad Man-Qi Ruan $^{2;2}$\email{ruanmq@ihep.ac.cn}%
\quad Da-Yong Wang $^{1;3}$\email{dayong.wang@pku.edu.cn}                                    \\
\quad Gang Li $^{2}$
\quad Shan Jin $^{2}$%\email{jins@ihep.ac.cn}%
\quad Yong Ban $^{1}$%\email{bany@pku.edu.cn}%
}
\maketitle

\address{%
$^1$ State Key Laboratory of Nuclear Physics and Technology, Peking University, Beijing 100871, China\\
$^2$ Institute of High Energy Physics, Chinese Academy of Sciences, Beijing 100049, China\\
}

\begin{abstract}
The Circular Electron Positron Collider (CEPC) is a future Higgs factory proposed by the Chinese high energy physics community. It will operate at a center-of-mass energy of 240-250 GeV. The CEPC will accumulate an integrated luminosity of 5 ab$^{\rm{-1}}$ in ten years' operation, producing one million Higgs bosons via the Higgsstrahlung and vector boson fusion processes. This sample allows a percent or even sub-percent level determination of the Higgs boson couplings. With GEANT4-based full simulation and dedicated fast simulation tool, we evaluated the statistical precisions of the Higgstrahlung cross section $\sigma_{ZH}$ and the Higgs mass $m_{H}$ measurement at the CEPC in the $Z\rightarrow\mu^+\mu^-$ channel. The statistical precision of $\sigma_{ZH}$ ($m_{H}$) measurement could reach 0.97\% (6.9 MeV) in the model-independent analysis which uses only the information of Z boson decay. For the standard model Higgs boson, the $m_{H}$ precision could be improved to 5.4 MeV by including the information of Higgs decays. Impact of the TPC size to these measurements is investigated. In addition, we studied the prospect of measuring the Higgs boson decaying into invisible final states at the CEPC. With the standard model $ZH$ production rate, the upper limit of ${\cal B}(H\rightarrow \rm{inv.})$ could reach 1.2\% at 95\% confidence level.
\end{abstract}



\footnotetext[0]{\hspace*{-3mm}\raisebox{0.3ex}{$\scriptstyle\copyright$}2013
Chinese Physical Society and the Institute of High Energy Physics
of the Chinese Academy of Sciences and the Institute
of Modern Physics of the Chinese Academy of Sciences and IOP Publishing Ltd}%

\begin{multicols}{2}

\section{Introduction}

The Higgs boson has been extensively studied since its discovery~\cite{:2012gk,:2012gu} at the LHC. The up-to-date results indicate that it is highly Standerd Model (SM) like~\cite{cmshig,lhcsub1,lhcsub2,lhcsub3,lhcsub4,lhcsub5}. On the other hand, many new physics models predict the Higgs couplings deviate from the SM at the percent level. Thus the
percent or even sub-percent level precision becomes necessary for the future Higgs measurement program. However, this accuracy is difficult to achieve at the LHC~\cite{hllhc}. Moreover, as the Higgs boson can only be reconstructed through its decay products at the LHC and thus it is impossible for the LHC to access the Higgs total width or absolute couplings in a model-independent way.

Compared to the hadron collider, an electron positron collider has significant advantages in the precision measurements of the Higgs boson. The beam energy and polarization of the initial states are precisely known and adjustable. Thus the Higgs production cross section is available with the recoil technique. In this way, a lepton collider can provide the absolute measurements of Higgs couplings~\cite{ILC,ilcsum,CEPC}. Besides, it is free of the QCD backgrounds. Almost every Higgs event can be recorded and reconstructed. Therefore, an electron-positron Higgs factory is an essential step in understanding the nature of the Higgs boson.

The Circular Electron Positron Collider is a Higgs factory proposed by the Chinese high energy physics community~\cite{CEPC}.
It will operate at a center-of-mass energy of 240-250 GeV with an instantaneous luminosity of 2 $\times$ $\rm{10}^{\rm{34}}$ $\rm{cm}^{\rm{-2}}$ $\rm{s}^{\rm{-1}}$. With two detectors operating over 10 years, the CEPC will accumulate about one million Higgs events, corresponding to an integrated luminosity of 5 ab$^{\rm{-1}}$.

The SM Higgs bosons are produced via the processes of $e^{+}e^{-}\rightarrow ZH$ (Higgsstrahlung),
$e^{+}e^{-}\rightarrow \nu\bar{\nu}H$ ($WW$ fusion) and $e^{+}e^{-}\rightarrow e^{+}e^{-}H$ ($ZZ$ fusion) at the CEPC~\cite{ww1,ww2,ww3,ww4,ww5,ww6}, as shown in Fig.~\ref{ZH}.
The corresponding production cross sections for the SM Higgs boson of 125 GeV,
as functions of center-of-mass energy, are plotted in Fig.~\ref{CrossSections}.
At the center-of-mass energy of 250 GeV, the Higgs bosons are dominantly produced from $ZH$ process,
where the Higgs boson is produced in association with a $Z$ boson.

\begin{center}
\centering
\includegraphics[width=7.5cm]{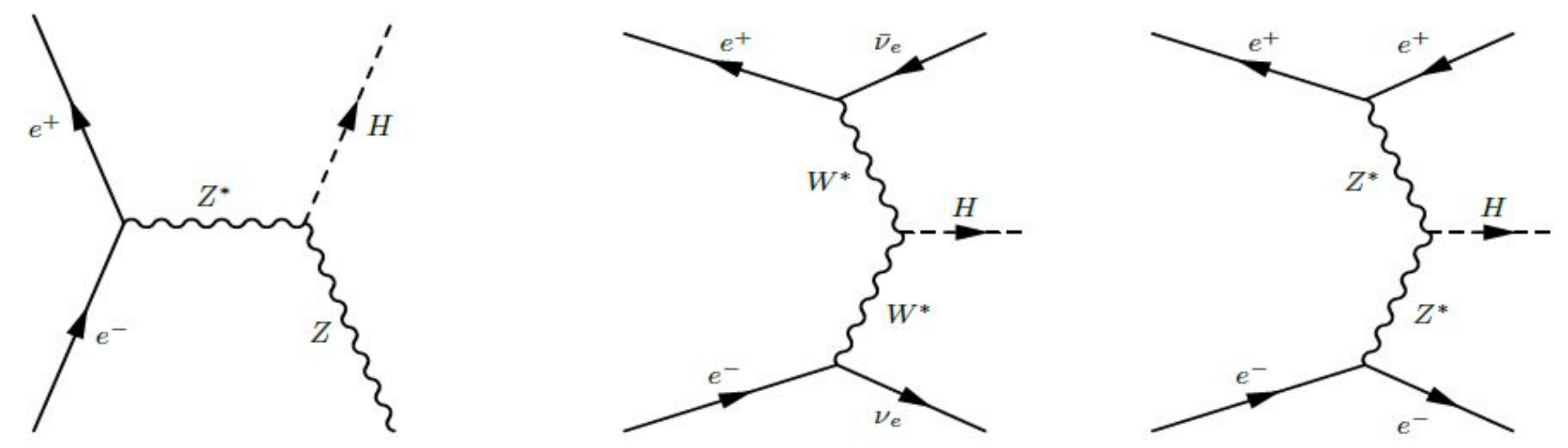}
\figcaption{ Feynman diagrams of the Higgs production mechanisms at the
  CEPC: the Higgsstrahlung, $WW$ fusion, and $ZZ$ fusion processes.}
\label{ZH}
\end{center}

\begin{center}
\centering
\includegraphics[width=7.5cm]{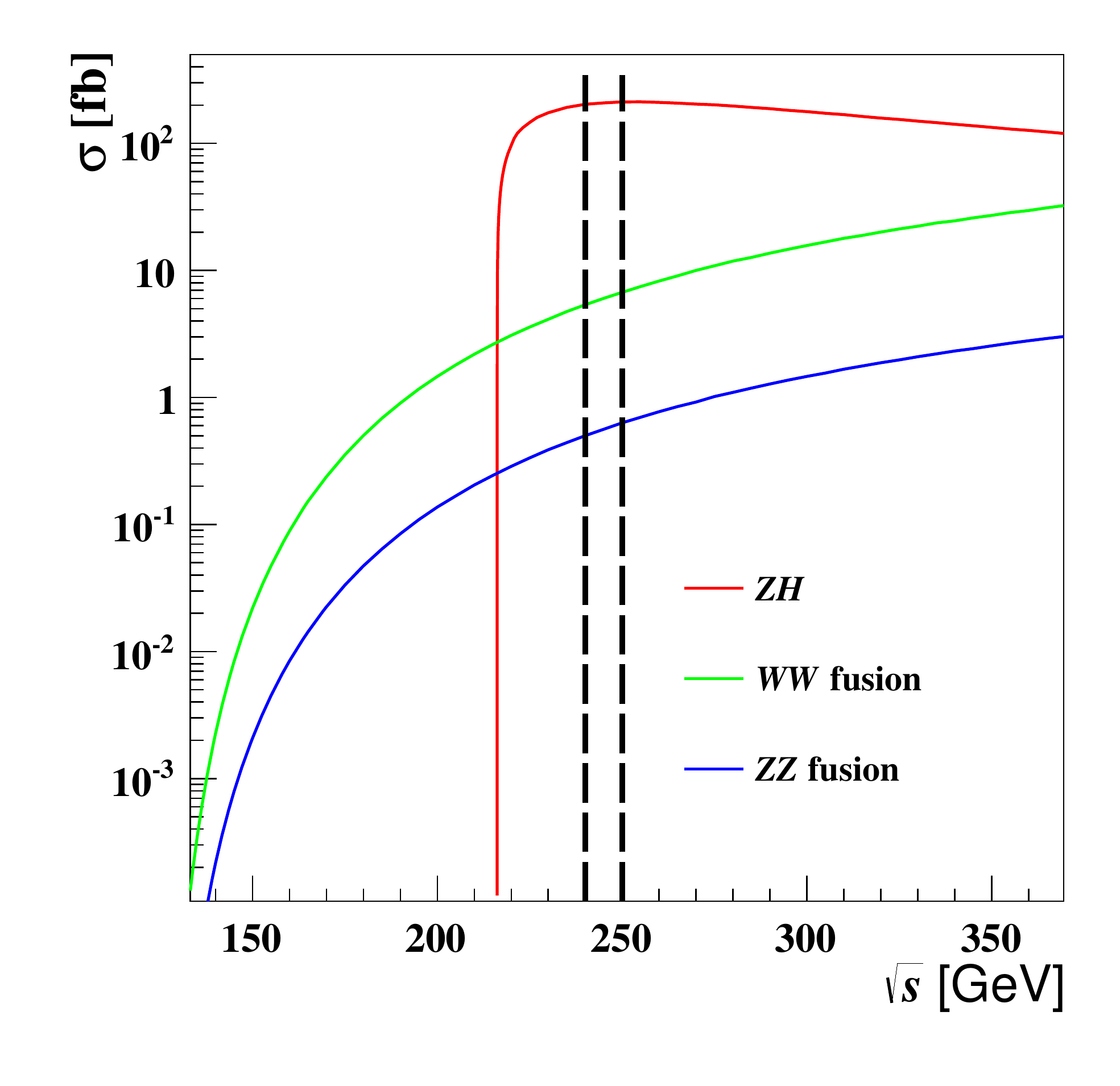}
\figcaption{Production cross sections of the Higgsstrahlung, $WW$ fusion
  and $ZZ$ fusion processes as functions of center-of-mass energy. The
  dashed lines (black) give the possible working energy range of the
  CEPC. }
\label{CrossSections}
\end{center}

The branching ratio of the $Z$ boson decaying into a pair of muons is 3.3\%.
The muons can be easily identified and their momentum can be precisely measured in the detector.
By tagging the muon pairs from the associated Z boson decays, the Higgsstrahlung events can be reconstructed with the recoil mass method:

~\\
\centerline{$M_{\rm{recoil}}= \sqrt{\emph{s} +M_{\mu^{+}\mu^{-}}^{2} - 2(E_{\mu^{+}} + E_{\mu^{-}})\sqrt{\emph{s}}}$~,}
~\\
where $E_{\mu^{+}}$ and $E_{\mu^{-}}$ are the energies of the two muons,
$M_{\mu^{+}\mu^{-}}$ is their invariant mass, and $s$ is the square of center-of-mass energy.
Therefore, the $ZH$ ($Z\rightarrow \mu^{+}\mu^{-}$) events form a peak in the $M_{\rm{recoil}}$ distribution at the Higgs boson mass.
%A typical $M_{\rm{recoil}}$ distribution of the Higgs signal is shown in Fig.~\ref{fastsimu}.
%The high mass tail is induced by radiation effects, including beamstrahlung, ISR, FSR, and bremstrahlung.

With the recoil mass method, the $ZH$ events are selected without using the decay information of the Higgs boson.
Thus the inclusive $ZH$ cross section $\sigma_{ZH}$ and the coupling $g_{HZZ}$ can be determined in a model-independent manner.
The measured $g_{HZZ}$, combined with exclusive Higgs boson decay measurements,
could be used to determine the Higgs boson width and
absolute values of couplings between the Higgs boson and its decay final states~\cite{ilcwidth}.
Meanwhile, the Higgs mass $m_{H}$ can be extracted from the $M_{\rm{recoil}}$ distribution.
A good knowledge of the Higgs mass is crucial since the $m_{H}$ is the only free parameter in the SM Higgs potential
and it determines the Higgs decay branching ratios in the SM.
Based on the model-independent analysis, the Higgs decay information can be used to further suppress the backgrounds, leading to a better $m_{H}$ precision.

The recoil mass method allows better exclusive measurement of Higgs decay channels.
Many new physics models predict a significant branching ratio of the Higgs boson decaying to invisible products~\cite{inv3,inv4,inv5,inv6}.
At the LHC, the current upper limit of this branching ratio is about 40\%~\cite{inv1,inv2},
which is much larger than the value predicted in the SM
(${\cal B}{(H\rightarrow \rm{inv.})}={\cal B}{(H\rightarrow ZZ\rightarrow \nu\nu\bar{\nu}\bar{\nu})}$ = 1.06$\times \rm{10}^{\rm{-3}}$).
At the CEPC, this measurement can be significantly improved by using the recoil mass method.
In this paper, we evaluate the upper limit on the branching ratio of the Higgs decaying to invisible final states.

A series of simulation studies of similar processes have been performed at the International Linear Collider (ILC)~\cite{ILC, ilcrecoil2}.
%A series of simulation researches have been performed on the future $e^+e^-$ colliders.
%The International Linear Collider (ILC) community has evaluated the inclusive Higgs production cross section and the Higgs mass using full detector simulation in the
%$Z\rightarrow e^+e^-$ and $Z\rightarrow \mu^{+}\mu^{-}$ channels.
%Using an integrated luminosity of 2 ab$^{\rm{-1}}$ with the beam polarization $P(e^{-},e^{+})=(-0.8,0.3)$ at 250 GeV,
%the ILC could achieve an accuracy of 0.8\% in the cross section and a precision of 14 MeV in the Higgs mass~\cite{ILC, ilcrecoil2}.
Compared to the ILC, the collision environment of the CEPC is significantly different.
The ILC uses polarized beams while the CEPC has no beam polarization.
Besides, the beam spot size of the CEPC at the interaction point (IP) is much larger
than that of the ILC, leading to a much weaker beamstrahlung effect and a narrower beam energy spread~\cite{beamstrahlung,CEPC,ILC}. The details of
parameter comparison are listed in Table~\ref{tab:parameters}~\cite{beamstrahlung}. Due to the above differences, the cross sections for both signal and backgrounds are different.
Therefore, it is necessary to perform the full detector simulation at the CEPC.

\begin{center}
%\centering  % ±í¾ÓÖÐ
  \tabcaption{Comparison of machine and beam parameters between the CEPC and the ILC.}
  \footnotesize
  \begin{tabular*}{80mm}{@{\extracolsep{\fill}}ccc}  \toprule
  Parameters                                                        &CEPC                        &ILC                          \\  \hline
%\hline
  Horizontal beam size at IP                                        &73700 nm                    &729 nm                       \\
  Vertical beam size at IP                                          &160 nm                      &7.7 nm                       \\
  Beamstrahlung parameter                                           &4.7$\times$10$^{\rm{-4}}$   &2.0$\times$10$^{\rm{-2}}$    \\
  Beam energy spread                                                &0.16\%                      &0.24\%                       \\
  Integrated luminosity                                             &5 ab$^{\rm{-1}}$            &2 ab$^{\rm{-1}}$             \\ \bottomrule
\end{tabular*}
\label{tab:parameters}
\end{center}

This paper is organized as follows. Section 2 describes the detector model,
Monte Carlo (MC) simulation and samples used in the studies. Section 3.1 presents the measurements of $ZH$ cross section and
the Higgs mass in a model-independent manner. The dependencies of measurement precisions on the TPC radii are investigated in Section 3.2, providing
a reference for the future detector optimization. The model-dependent analysis of the Higgs boson mass is described in Section 3.3, and the
measurement of the Higgs decaying to invisible final states is presented in Section 3.4.
In Section 4, we discuss the systematic uncertainties and the methodology of systematic control.
The conclusion is summarized in section 5.

\section{Detector and Monte Carlo Simulation}

The analysis is performed on the MC samples simulated on the CEPC conceptual detector,
which is based on the International Large Detector (ILD)~\cite{ild1,ild2} at the ILC~\cite{ILC}.
With respect to the ILD, the CEPC conceptual detector has a $L^{*}$
(the distance between the interaction point and QD0, the final focusing magnet) of 1.5 m, which is significantly shorter than that of the ILC (4.5 m).
The shorter $L^{*}$ is essential to achieve a high luminosity by reducing the beam nonlinearity in the interaction region. Besides,
the CEPC has multiple interaction points, thus the push-pull operation is not necessary.
Therefore, the thickness of return Yoke is reduced by 1 meter at the CEPC conceptual detector.

Apart from the $L^{*}$ and return Yoke, the CEPC conceptual detector follows the same design of the ILD.
Installed in a solenoidal magnet of 3.5 Tesla, the CEPC conceptual detector consists of a vertex detector, a tracking system
and a calorimetry system. The silicon pixel vertex detector (VTX) consists of three cylindrical and concentric double-layers, with an innermost radius of 16 mm~\cite{ild1,ild2}.
The tracking system is composed of a time projection chamber (TPC) as the main tracker and the silicon tracking devices, including
a silicon inner tracker (SIT), forward tracking disks (FTDs), a silicon external tracker (SET) and
end-cap tracking disks. The VTX and SIT are expected to provide a spatial resolution of better than 3 $\mu$m near the interaction point,
which is crucial for the vertex reconstruction and the jet flavor tagging.
The outermost FTD disk is positioned at $z=$ 1057.5 mm to the IP. With an inner radius of 92.7 mm and an outer radius of 309 mm, it improves the geometric acceptance of
the tracking system to $|cos\theta|<$0.995.
The TPC has nearly 200 three-dimensional ($r$, $\phi$ and $z$) spacepoints, with inner and outer radii of 0.325 m and 1.8 m respectively, and a half-length of 2.35 m. It provides an expected spatial resolution of better than 100 $\mu$m in the $r\phi$ plane.
The SET provides a precise position measurement outside the TPC.
Such a tracking system is expected to achieve a precise determination of the charged particle momenta with a resolution of
$\sigma_{1/p_{T}}=2\times 10^{-5}\bigoplus 1\times \frac{10^{-3}}{p_{T}\cdot sin\theta}$, where $p_{T}$ is
the transverse momentum and $\theta$ is the polar angle.
The calorimetry system is composed of a high granularity electromagnetic calorimeter and a high granularity hadron calorimeter,
allowing excellent separations of showers of different particles. It is expected to provide a jet energy resolution of 3-4\% and enable
the PID efficiency to be over 99.5\% for muons with momentum larger than 10 GeV. More information about the CEPC conceptual detector can be found
in reference~\cite{CEPC}.

A set of event samples at a center-of-mass energy of 250 GeV,
corresponding to an integrated luminosity of 5 ab$^{\rm{-1}}$, has been generated with Whizard 1.95~\cite{whizard1,whizard2}.
It consists of the SM Higgs signal with $m_{H}$ = 125 GeV and the major SM backgrounds,
including the $\gamma\gamma$ process (photon-induced background $e^+e^-\rightarrow e^+e^-\gamma\gamma\rightarrow e^+e^-l^+l^-$, where the photons are
generated according to the Weizs\"{a}cker-Williams approximation~\cite{wwa1,wwa2,wwa3}), 2-fermion processes ($e^+e^-\rightarrow f\bar{f}$,
where $f\bar{f}$ refers to all lepton and quark pairs)
and 4-fermion processes, categorized as $ZZ$, $WW$, $ZZ$ or $WW$, single $Z$ ($e^+e^- Z$) and single $W$ ($e^+\nu_e W$ or $e^-\bar{\nu_e} W$).
If the final states could be produced through both $WW$ and $ZZ$ intermediate states, such as $e^{+}e^{-}\nu_{e}\bar{\nu_{e}}$, this process is classified as ``$ZZ$ or $WW$''
and their  interference is included.
%If the final state is $e^{+}e^{-}$ together with another pair of lepton, it is classified into the single Z category no matter it is from intermediate Z boson or photon.
The initial state radiation (ISR) is also taken into account in the sample generation.
%The $ZZ$ and $WW$ processes correspond to $e^{+}e^{-}\rightarrow ZZ/WW$ where final states fermions has no ambig.
%Some final states consist of two mutually charge conjugated fermion pairs, which could be from both $WW$ and $ZZ$, and this type is called ``$ZZ$ or $WW$''.
More details about the CEPC samples can be found in reference~\cite{samples}.

The Higgs signal samples are fully simulated with Mokka~\cite{Mokka} and reconstructed with ArborPFA~\cite{arbor}.
A beam energy spread of 0.16\% has been included in this analysis.
In order to save the computing power, a fast simulation framework has been developed to process the backgrounds.
In the fast simulation, the detector responses are obtained by a series of full simulations for single particle events. Then the responses,
including momentum resolution and detection efficiency, have been parameterized as functions of
energy and polar angle for different types of particles. The four-momenta of the visible final state particles are smeared according to
the parameterized resolutions and they are randomly accepted based on the corresponding detection efficiencies.
%A validation between the fast and full simulation with the $ZZ$ events is shown in Fig.~\ref{fastsimu}.

\section{The Analyses}

The expected number of $ZH$ events $N_{ZH}$ can be expressed as
\begin{equation}
N_{ZH}= \sigma_{ZH}\cdot {\cal L} \cdot \sum_{X} \epsilon(H\rightarrow X)\cdot {\cal B}(H\rightarrow X)~,
\end{equation}
where $\cal{L}$ is the integrated luminosity,
${\cal B}(H\rightarrow X)$ is the branching ratio of an exclusive Higgs decay mode, and $\epsilon(H\rightarrow X)$ is the corresponding selection efficiency.
In the model-independent analysis using only information in the Z boson decays, the efficiencies are expected to be
uniform for each Higgs decay mode, we can write
\begin{equation}
N_{ZH}=\sigma_{ZH}\cdot {\cal L} \cdot \epsilon \cdot \sum_{X} {\cal B}(H\rightarrow X)=\sigma_{ZH}\cdot {\cal L} \cdot \epsilon~,
\end{equation}
and thus $\sigma_{ZH}$ can be determined in a model-independent manner.
The Higgs decay information can be used to further suppress the SM backgrounds and improve the precision
on the mass measurement.
In this case, the selection efficiency $\epsilon(H\rightarrow X)$ depends on the Higgs decay mode.
%and $\sigma_{ZH}$ can not be factorized directly from the $N_{ZH}$ measurement.

\subsection{Model-independent analysis on $\sigma_{ZH}$ and $m_{H}$ measurement}

In the model-independent analysis, the event selection is composed of a pre-selection and a multivariate analysis (MVA).
In the pre-selection, a pair of oppositely charged muons is required.
The pair with the minimum $|M_{\mu^{+}\mu^{-}}-M_{Z}|$ is selected in case of multi-combinations, where $M_{Z}$ is 91.2 GeV~\cite{pdg}.
The invariant mass of $\mu^{+}\mu^{-}$ is required to satisfy 80 GeV $<\ M_{\mu^{+}\mu^{-}}\ <$ 100 GeV.
In order to suppress 2-fermion backgrounds, the transverse momentum of the muon pair, $p_{\rm{T}\mu^{+}\mu^{-}}$, is required to be larger than 20 GeV and
the difference of the azimuth angles of the two muons should be less than $\rm{175}^{\circ}$.
\iffalse
In the pre-selection, the selection criteria are listed below.
\begin{enumerate}[(1)]
\item At least one pair of muons with opposite charge are reconstructed. For events with more than one combinations, the one with the minimum $|M_{\mu^{+}\mu^{-}}-M_{Z}|$ is selected, where $M_{\mu^{+}\mu^{-}}$ is the invariant mass of the reconstructed $Z$ boson and $M_{Z}$ is 91.2 GeV~\cite{pdg}.

\item The invariant mass of $\mu^{+}\mu^{-}$ is required to satisfy 80 GeV $<\ M_{\mu^{+}\mu^{-}}\ <$ 100 GeV, while the recoil mass is in the region of 120 GeV $<\ M_{recoil}\ <$ 160 GeV.

\item The transverse momentum $p_{\rm{T}\mu^{+}\mu^{-}}$ is required to be larger than 20 GeV. In addition, it is required that $|\phi_{\mu^{+}}-\phi_{\mu^{-}}|\ <\ \rm{175}^{\circ}$, where $\phi_{\mu^{\pm}}$ is the azimuth angle of $\mu^{+}$ or $\mu^{-}$.

\end{enumerate}
\fi
The Toolkit for Multivariate Analysis (TMVA)~\cite{TMVA} is used to the further background rejection.
In this paper, the method of gradient Boosted Decision Trees (BDT) is adopted and the
selected variables for TMVA input are $M_{\mu^{+}\mu^{-}}$,
$p_{\rm{T}\mu^{+}\mu^{-}}$, the polar angle of Z candidate, and the
acollinearity of muon pair, which is defined as

\begin{equation}
acol=\rm{cos}^{\rm{-1}}\frac{\textbf{\emph{p}}_{\mu^{+}}\cdot \textbf{\emph{p}}_{\mu^{-}}}{|\textbf{\emph{p}}_{\mu^{+}}|\cdot|\textbf{\emph{p}}_{\mu^{+}}|}~,
\end{equation}
where $\textbf{\emph{p}}$$_{\mu^{\pm}}$ is the momentum vector of $\mu^{\pm}$. After the pre-selection, a half of the remaining backgrounds are selected
for training, together with another copy of signal sample of 5 $ab^{\rm{-1}}$. The BDT response is calculated using weights obtained from
training samples and applied to the whole data set, shown in Fig.~\ref{BDT}. With the requirement of BDT$>$-0.05, the signal/background ratio is improved from
12.3\% to 31.1\%.

The BDT selection is optimized to the $\sigma_{ZH}$ measurement. The cut flow is summarized in Table~\ref{tab:cutflow} and the signal selection efficiency is 62.8\%.
After the selection, the leading backgrounds are from $ZZ$ (18.8\% of the remaining background), $\gamma\gamma$ (21.8\%) and 2-fermion (32.8\%) processes.
The selected muons may also come from the ZH events with the Z boson not decaying to $\mu^{+}\mu^{-}$. About 200 events of this type survive after the event selection
and they are flatly distributed in the signal region. They are neglected due to their small contribution ($\sim$ 0.29\%) in the total background.

\begin{center}
\centering
\includegraphics[width=7.0cm]{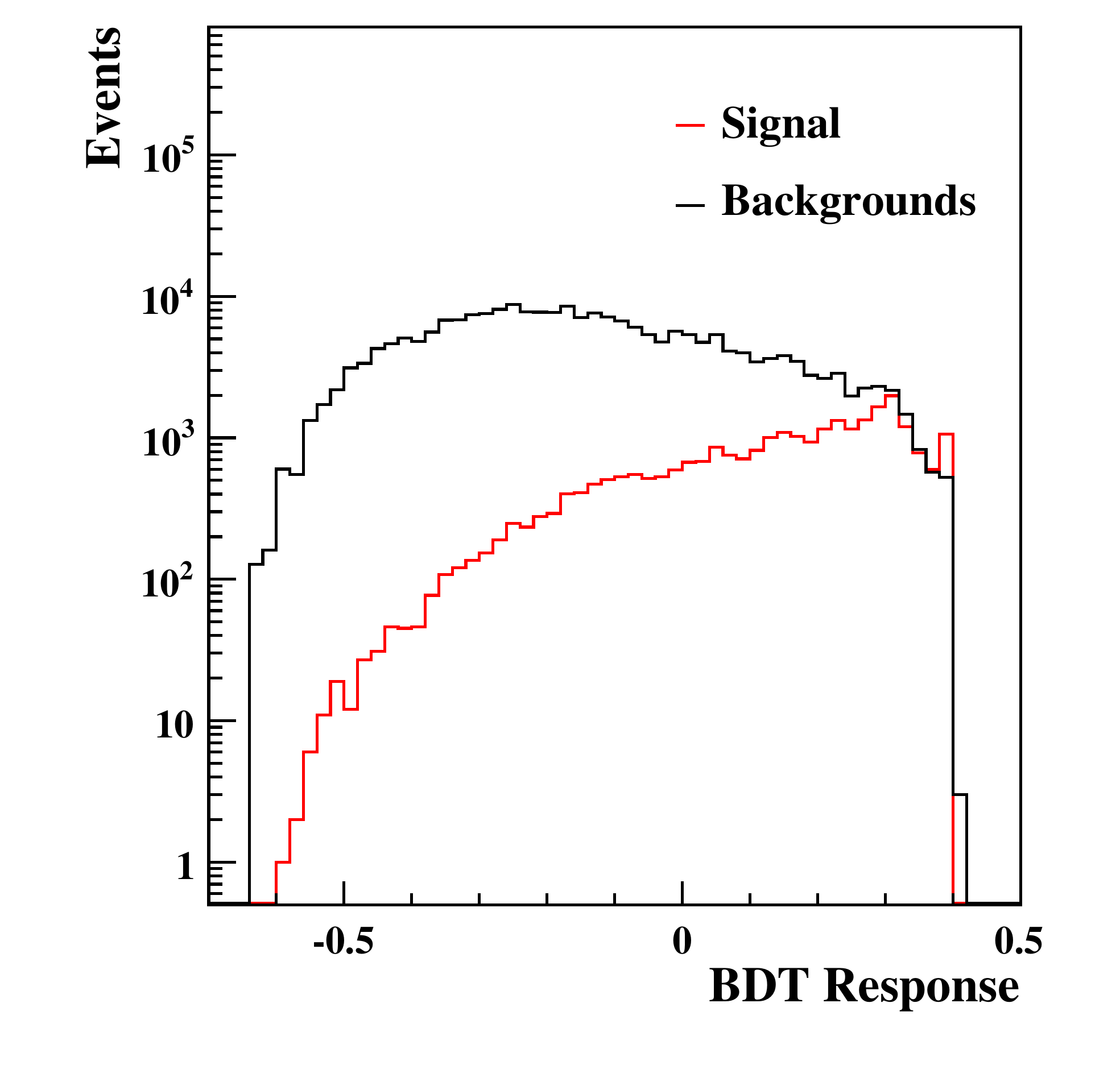}
\figcaption{The BDT response for the signal and background samples. The red solid line is signal and the black
dashed line is background. The number of background is normalized to that of signal. }
\label{BDT}
\end{center}

\end{multicols}

\begin{center}
%\centering
  \tabcaption{Efficiencies of signal and background in the model-independent analysis}
  \footnotesize
  \begin{tabular*}{170mm}{@{\extracolsep{\fill}}cccccccc}  \toprule
%\begin{tabular}{lcccccccccc}
                                                                    &$Z(\mu^+\mu^-)H$ &$ZZ$    &$WW$     &$ZZ$ or $WW$ &Single $Z$  &$Z$(2$f$)  &$\gamma\gamma$  \\  \hline
  total generated                                                   &35247            &5347053 &44180832 &17801222     &7809747     &418595861  &161925000       \\
  $N_{\mu^{+}}$ $\ge$ 1, $N_{\mu^{-}}$ $\ge$ 1                      &95.7\%           &11.95\% &0.65\%   &3.92\%       &9.75\%      &1.64\%     &17.31\%         \\
  120 GeV $<\ M_{recoil}\ <$ 150 GeV                                &93.2\%           &1.71\%  &0.23\%   &0.70\%       &1.93\%      &0.17\%     &3.06\%          \\
  80 GeV $<\ M_{\mu^{+}\mu^{-}}\ <$ 100 GeV                         &85.5\%           &0.68\%  &0.06\%   &0.22\%       &0.22\%      &0.10\%     &0.11\%          \\
  $p_{\rm{T}\mu^{+}\mu^{-}}\ >$ 20 GeV                              &80.2\%           &0.57\%  &0.06\%   &0.17\%       &0.16\%      &0.02\%     &0.04\%          \\
  $\Delta\phi\ <$ 175$^{\circ}$                                     &77.8\%           &0.51\%  &0.05\%   &0.17\%       &0.15\%      &0.01\%     &0.04\%          \\
  BDT cut                                                           &63.0\%           &0.25\%  &0.01\%   &0.05\%       &0.06\%      &0.01\%     &0.01\%          \\
  fit window                                                        &62.8\%           &0.25\%  &0.01\%   &0.05\%       &0.05\%      &0.01\%     &0.01\%          \\  \bottomrule
\end{tabular*}
\label{tab:cutflow}
\end{center}

\begin{multicols}{2}

The final recoil mass spectrum of $\mu^{+}\mu^{-}$ is shown in Fig.~\ref{recoilmi}.
An unbinned maximum likelihood fit to the $M_{recoil}$ distribution is performed in the region of 120 GeV to 140 GeV to determine the signal yield as well as the value of the Higgs mass.
The background is represented by a third order Chebychev polynomial function, whose parameters are fixed to the values extracted from the background samples.
The Higgs signal shape is described by a Crystal Ball function.
Based on the fit results, $\sigma_{ZH}$ is estimated to a relative precision of 0.97\% and $m_{H}$ to a precision of 6.9 MeV.

\begin{center}
\centering
\includegraphics[width=7.0cm]{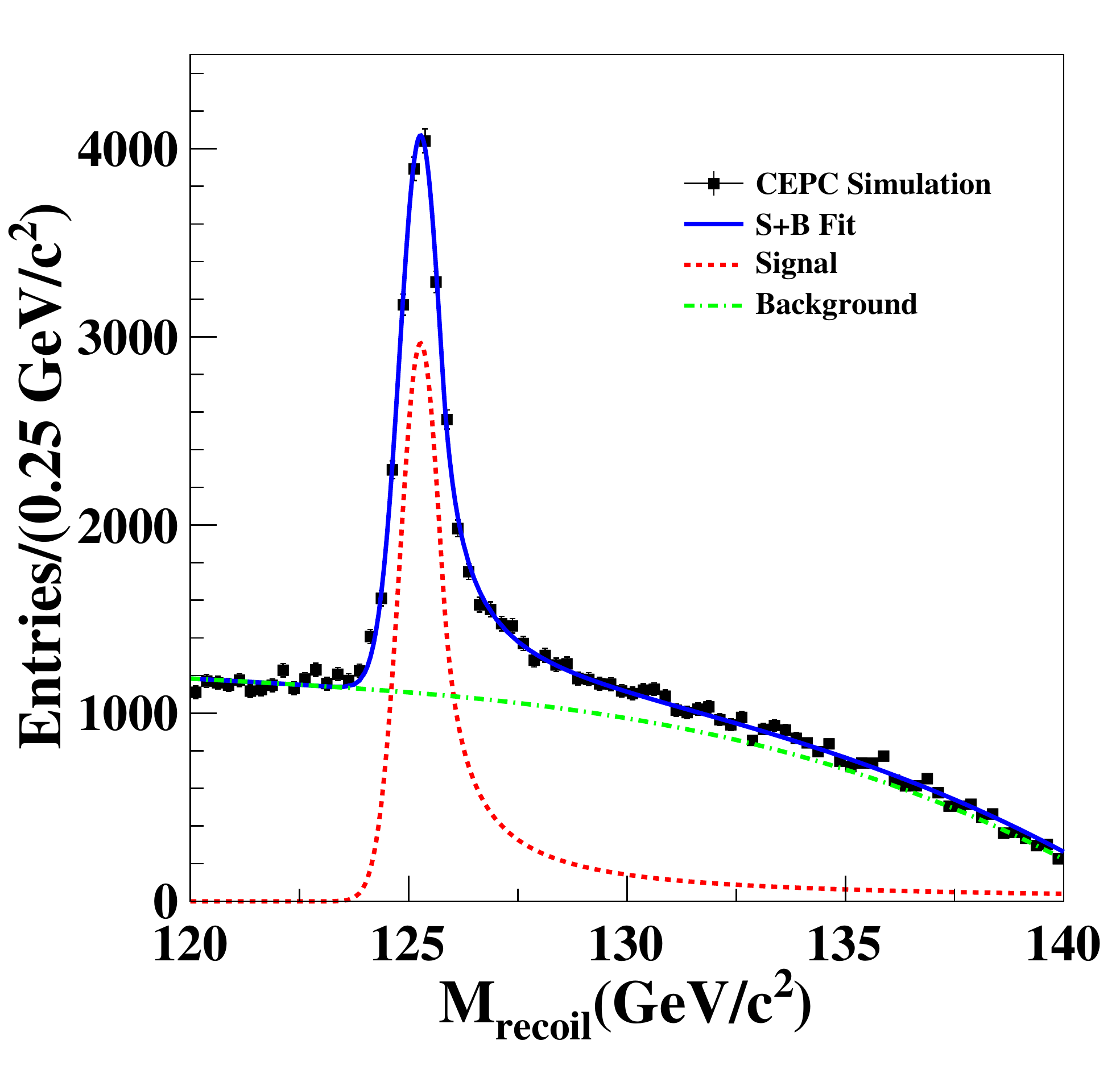}
\figcaption{The recoil mass spectrum of $\mu^{+}\mu^{-}$ in the model-independent analysis.
  The dots with error bars represent the CEPC simulation.
  The solid (blue) line indicates the fit.
  The dashed (red) and the long-dashed (green) line show the signal and the background contributions in the fit, respectively. }
\label{recoilmi}
\end{center}

The uniformity of event section efficiency with different Higgs decay modes is studied.
A SM $ZH\ (Z\rightarrow\mu^+\mu^-)$ sample corresponding to 500 $ab^{\rm{-1}}$ has been simulated, where the Higgs boson decays inclusively.
Fig.~\ref{effmi} shows the efficiencies and no significant bias to any specific Higgs decay mode is observed.
In order to evaluate the impact of the sensitivities to the various Higgs decay modes, %% ????
the SM $ZH$ cross section is kept unchanged and a specific SM Higgs decay branching ratio is enlarged by 5\% each time
(${\cal B}(H\rightarrow X)\rightarrow {\cal B}(H\rightarrow X)$ + 5\%)).  All branching ratios are then scaled to
keep the total event rate and the resultant differences in the $\sigma_{ZH}$ measurement are summarized in Table~\ref{tab:eff}.
The largest bias in the $\sigma_{ZH}$ is less than 10$^{\rm{-3}}$,
which is much smaller than the statistical uncertainty 0.97\%.
It is therefore reasonable to conclude the recoil mass method gives a model-independent measurement.

\begin{center}
\centering
\includegraphics[width=7.cm]{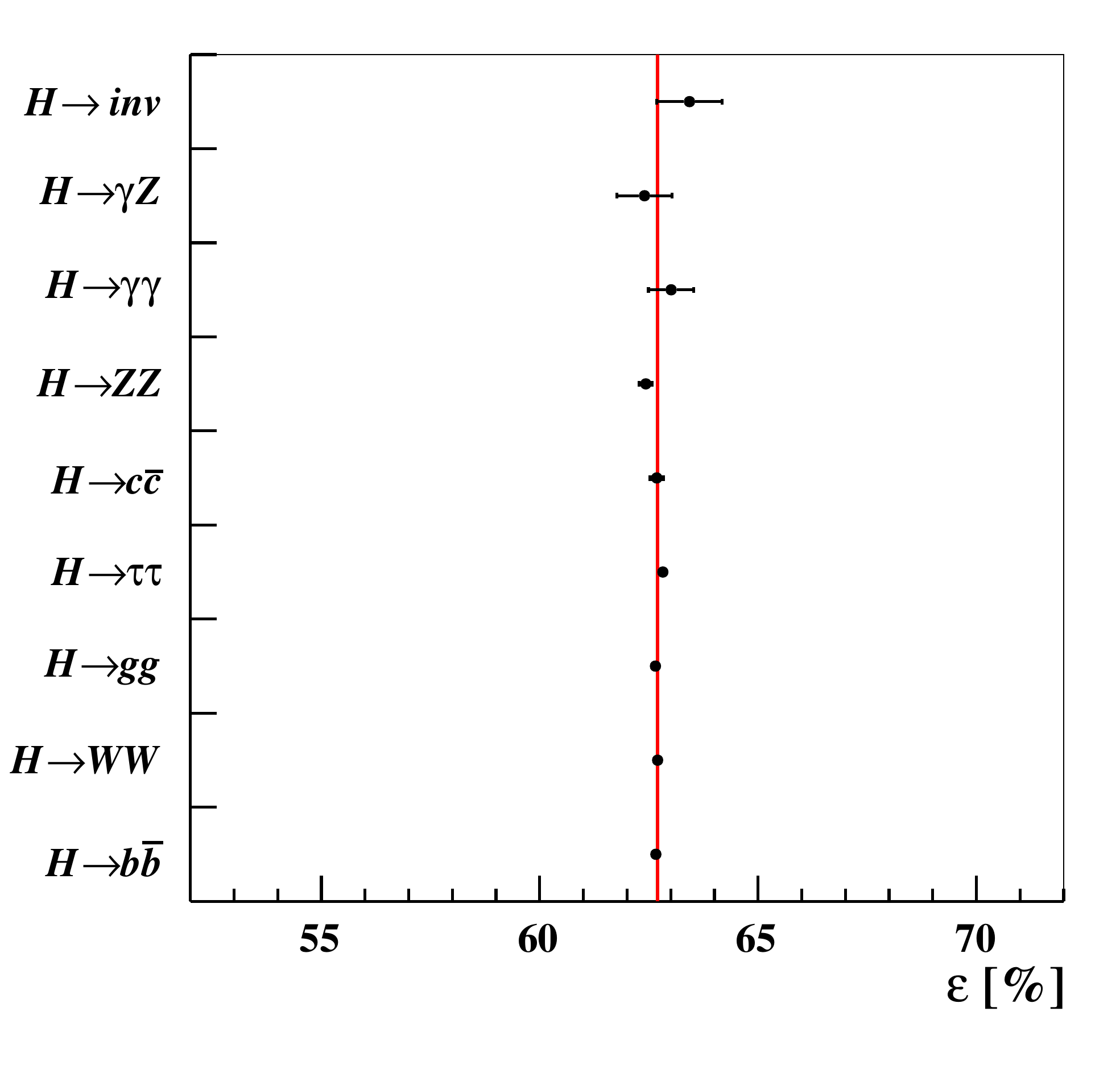}
\figcaption{The efficiencies for main decay modes of the
  Higgs boson in model-independent analysis. The dots with error bars are efficiencies from exclusive
  channels. The solid line (red) is the event selection efficiency of
  model-independent analysis.}% and the band (green) indicates the averaged one standard deviation. }
\label{effmi}
\end{center}

\begin{center}
%\centering
  \tabcaption{Estimation of biases of $\sigma_{ZH}$ caused by potential variances of the Higgs decay branching ratios.}
  \footnotesize
  \begin{tabular*}{50mm}{@{\extracolsep{\fill}}cc}  \toprule
  Decay mode                                                        &Bias($\times 10^{-4}$)                 \\  \hline
%\hline
  $H\rightarrow b\bar{b}$                                           &-0.10              \\
  $H\rightarrow WW$                                                 &+0.20              \\
  $H\rightarrow gg$                                                 &-0.18              \\
  $H\rightarrow \tau\tau$                                           &+1.11              \\
  $H\rightarrow c\bar{c}$                                           &+0.05              \\
  $H\rightarrow ZZ$                                                 &-1.85              \\
  $H\rightarrow \gamma\gamma$                                       &+2.56              \\
  $H\rightarrow \gamma Z$                                           &-2.08              \\
  $H\rightarrow$ inv.                                               &+5.75              \\  \bottomrule
\end{tabular*}
\label{tab:eff}
\end{center}

\subsection{The dependency of $\sigma_{ZH}$ and $m_{H}$ measurement accuracies on the TPC radius}

From the detector point of view, the precisions of $\sigma_{ZH}$ and $m_{H}$ are mainly determined
by the detector solid angle acceptance and the muon identification efficiency. Besides, the $m_{H}$ precision
also relies on the muon momentum resolution.
%The momentum resolution strongly depends on the tracker design especially the tracker radius.
%The tracking system of the CEPC conceptual detector is composed of a silicon tracking system and a main tracker of TPC.
The momentum resolution scales approximately with the inverse of $BL^2$, where $B$ and $L$ represent the magnetic filed strength and
the detector radius respectively. The tracking system of the CEPC conceptual detector is composed of a silicon tracking system and a
main tracker of TPC. A larger TPC radius, corresponding to a larger lever arm, will give a better momentum resolution but it brings more construction cost.
The performances at different TPC radii are studied using the fast simulation tool. In the model-independent analysis, the expected accuracies of
$\sigma_{ZH}$ and $m_{H}$ are recorded, see Fig.~\ref{tpc1}. If the TPC radius is reduced by 25\%,
the precisions of $\sigma_{ZH}$ and $m_{H}$ are worsened by 2\% and 20\%, respectively.
These expected accuracies are then parameterized as functions of the TPC radius. For the $\sigma_{ZH}$ measurement, it is expressed as
\begin{equation}
\frac{\delta\sigma_{ZH}}{\sigma_{ZH}}=\rm{0.52}\ \times\ (\rm{1} + \rm{e}^{\rm{-0.09}\cdot \emph{R}_{\rm{TPC}}})~,
\end{equation}
where $\frac{\delta\sigma_{ZH}}{\sigma_{ZH}}$ (\%) is the relative precision of cross section
measurement and $R_{\rm{TPC}}$ (m) is the TPC radius.
Similarly, the accuracies of Higgs mass $\delta m_{H}$ at different TPC radii are obtained.
Its dependence on the TPC radius can be expressed as
\begin{equation}
\delta m_{H}=5.85\ \times\ (\rm{1}+ \rm{5.19}\times \rm{e}^{\rm{-1.81}\cdot \emph{R}_{\rm{TPC}}})\ MeV.
\end{equation}

\begin{center}
\centering
\includegraphics[width=7.cm]{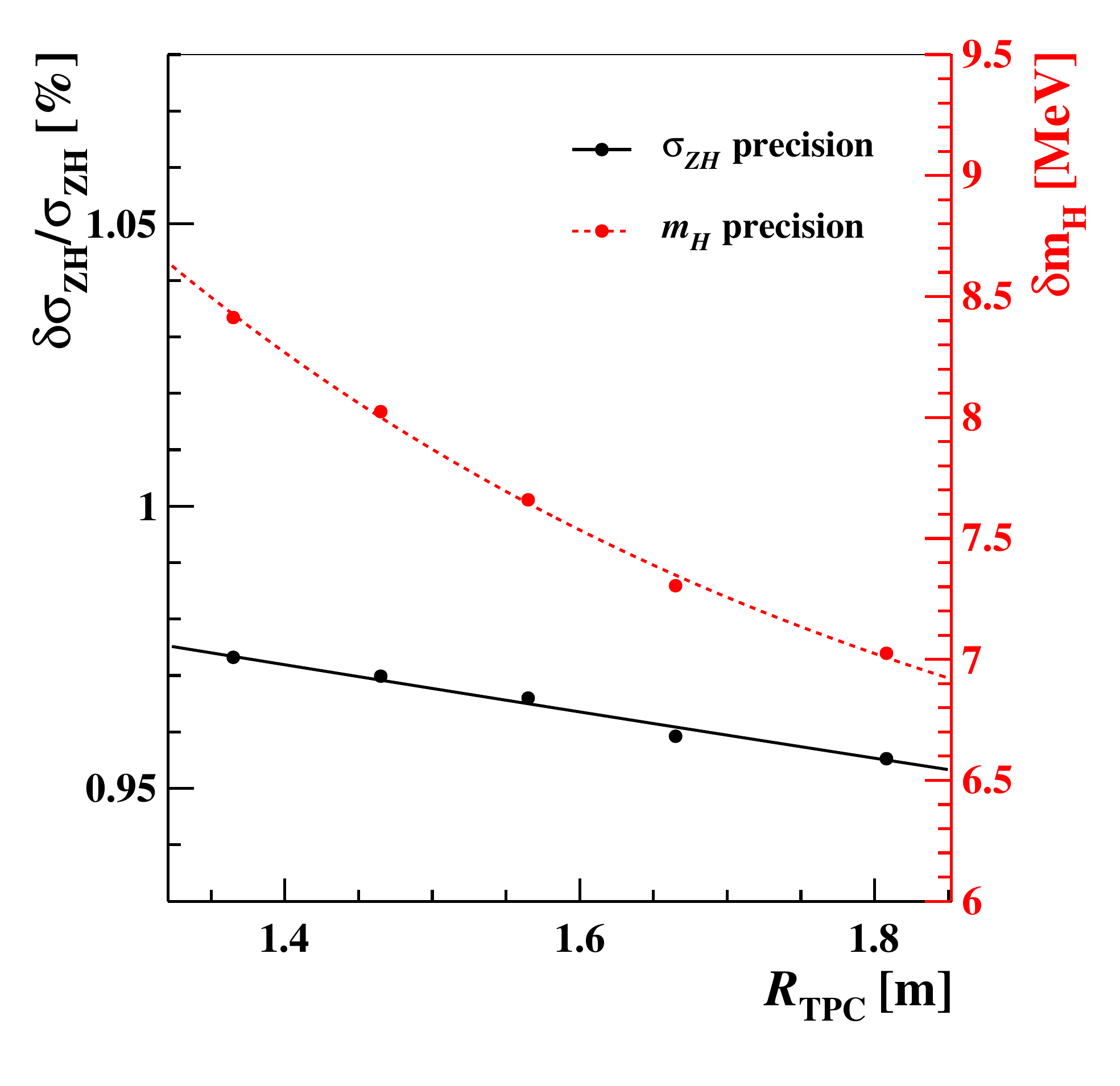}
\figcaption{The precisions of $\sigma_{ZH}$ and $m_{H}$ measurements
  versus different TPC radii. The solid line represents the precision of $\sigma_{ZH}$, and
  the dashed line is for $m_{H}$. }
\label{tpc1}
\end{center}

\subsection{Model-dependent analysis on $m_{H}$ measurement}

Assuming the SM Higgs decays, the background can be further suppressed by using the Higgs decay information, leading to
a better Higgs mass measurement.
On top of the pre-selection criteria used in the model-independent analysis,
we request that there are more than four charged tracks reconstructed.
In the MVA stage, the energy of all reconstructed final states, $E_{\rm{vis}}$, is also taken as an input variable except those in the MVA of model-independent analysis.
After the final selection, the recoil mass spectrum is shown in Fig.~\ref{recoilmd} and an efficiency of 66.1\% is obtained.
The resultant precision of $m_{H}$ is improved to 5.4 MeV.

\begin{center}
\centering
\includegraphics[width=7.cm]{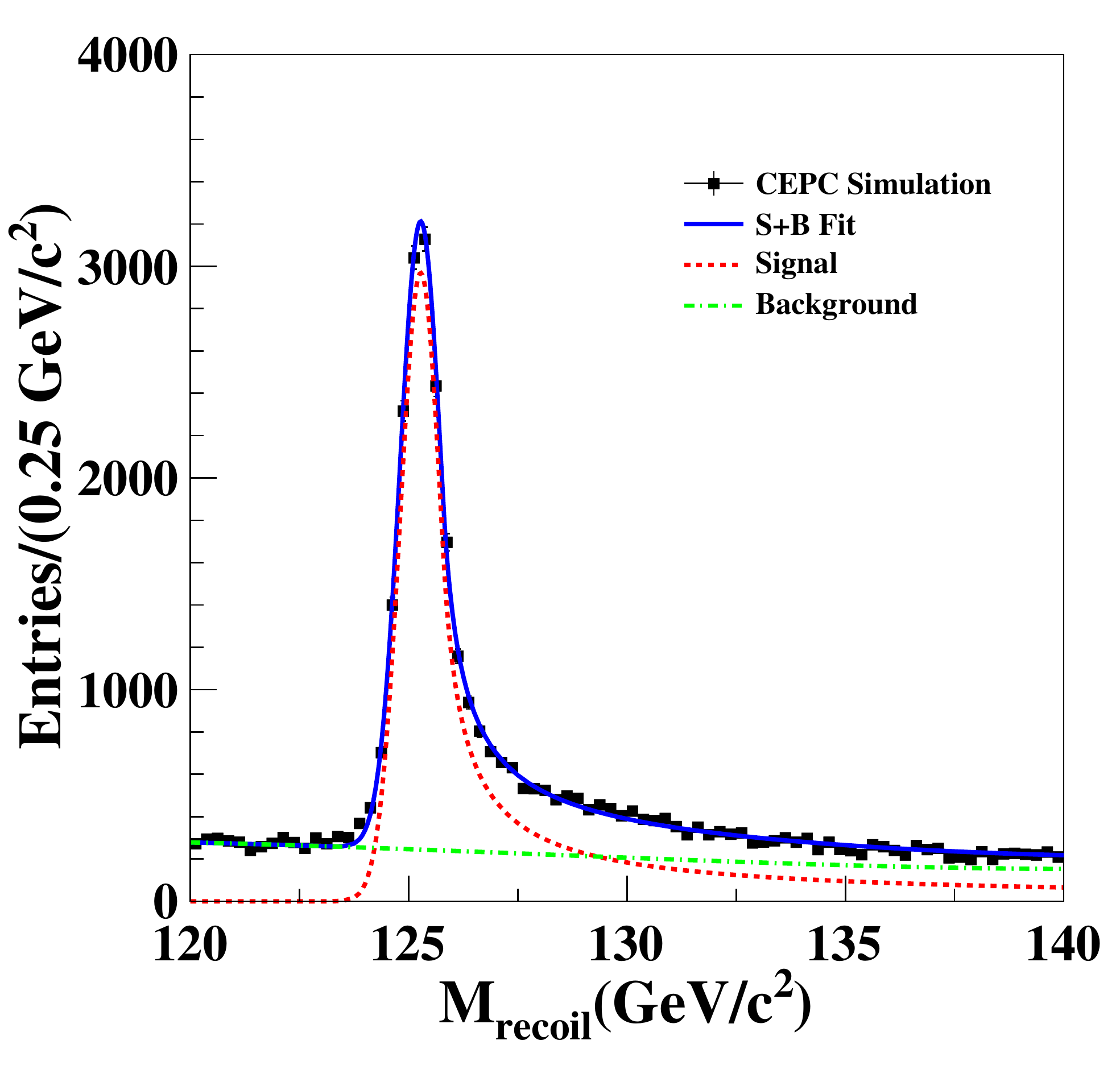}
\figcaption{The recoil mass spectrum of $\mu^{+}\mu^{-}$ in the model-dependent analysis.
  The dots with error bars represent the CEPC simulation.
  The solid (blue) line indicates the fit.
  The dashed (red) and the long-dashed (green) line are the signal and the background, respectively.
}
\label{recoilmd}
\end{center}

\subsection{The measurement of the invisible decay mode of the Higgs boson}

The invisible decay mode of the Higgs boson is a well motivated signature of the physics beyond the SM~\cite{inv3,inv4,inv5,inv6}.
At the CEPC, the invisible Higgs boson decay branching ratio can be determined precisely using the recoil mass method.
Assuming the Higgs boson has the SM coupling to the Z boson and non-vanishing couplings to the beyond SM invisible particles, the
measurement potential of the Higgs boson decaying to invisible final states at the CEPC is investigated at different values of
${\cal B}(H\rightarrow \rm{inv.})$.

The signal candidates are identified using the same pre-selection as that in the model-independent analysis.
In addition, we require that there is no extra visible energy except that of the muon pair decayed from $Z$ boson
and that $E_{\rm{vis}}$ must be within 105 GeV and 125 GeV.
%The requirement of MVA is optimized to achieve the best at various assumptions of ${\cal B}(H\rightarrow \rm{inv})$.
Fig.~\ref{recoilinv} shows the $\mu^{+}\mu^{-}$ recoil mass spectrum of the candidates with ${\cal B}(H\rightarrow \rm{inv.})$ = 50\%.
In this scenario, the final signal event selection efficiency is 63.9\% and the relative precision of the cross section of Higgs decaying to invisible
final states $\delta\sigma_{ZH, H\rightarrow \rm{inv.}}/\sigma_{ZH, H\rightarrow \rm{inv.}}$ reaches 1.16\%.

\begin{center}
\centering
\includegraphics[width=7.cm]{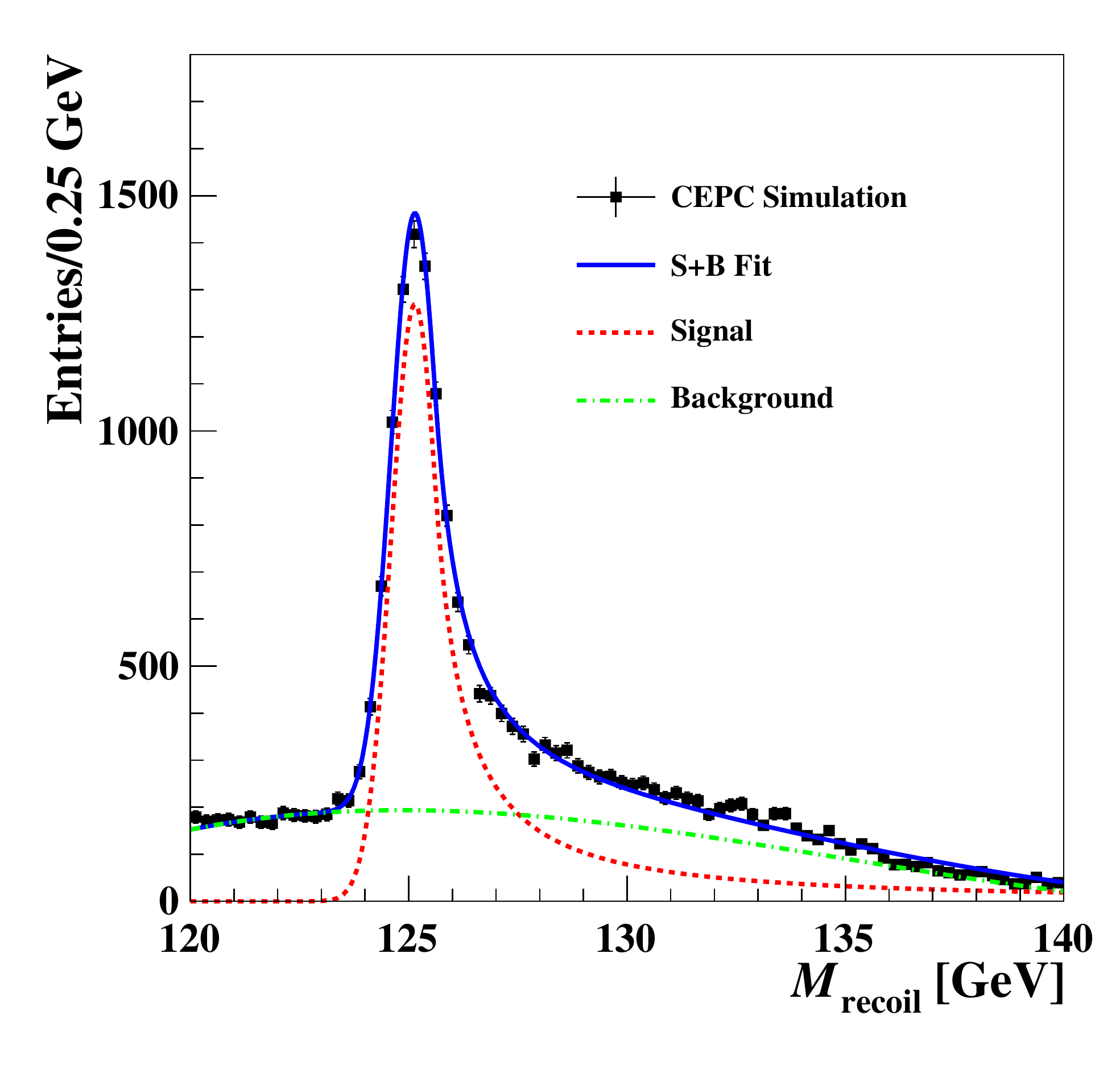}
\figcaption{The recoil mass spectrum of $\mu^{+}\mu^{-}$ in the measurement of the invisible decay mode of the Higgs boson with ${\cal B}(H\rightarrow \rm{inv.})$ = 50\%.
  The dots with error bars represent the CEPC simulation.
  The solid (blue) line indicates the fit.
  The dashed (red) and the long-dashed (green) line are the signal and the background, respectively.
}
\label{recoilinv}
\end{center}

\begin{center}
\centering
\includegraphics[width=6.7cm]{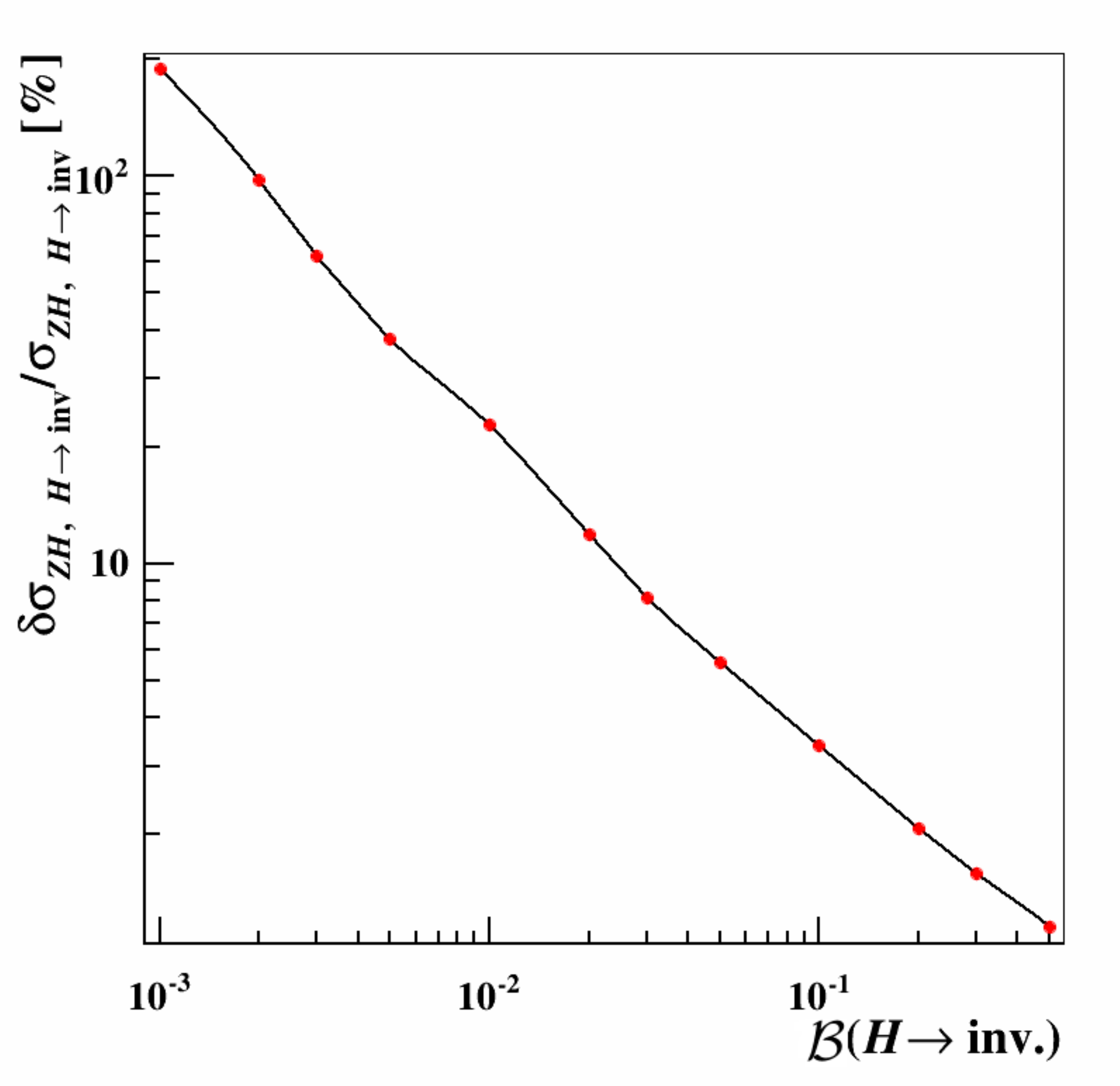}
\figcaption{The precision of the cross section of Higgs decaying to invisible final states
$\delta\sigma_{ZH, H\rightarrow \rm{inv.}}/\sigma_{ZH, H\rightarrow \rm{inv.}}$ versus ${\cal B}(H\rightarrow \rm{inv.})$.}
\label{invisible}
\end{center}

Based on different assumptions of ${\cal B}(H\rightarrow \rm{inv.})$,
the relative precisions of $\delta\sigma_{ZH, H\rightarrow \rm{inv.}}/\sigma_{ZH, H\rightarrow \rm{inv.}}$ are given in Fig.~\ref{invisible}.
The upper limit of ${\cal B}(H\rightarrow \rm{inv.})$ at 95\% confidence level is estimated to be 1.2 $\times$ 10$^{\rm{-2}}$ by using the likelihood ratio test method~\cite{likeli}.

\section{Discussion on systematic uncertainties}

A complete investigation of potential systematic uncertainties is beyond the scope of this paper.
Here we present several sources of systematic uncertainties and the strategies to deal with them. With
the best knowledge,

The common uncertainties on $\sigma_{ZH}$ and $m_{H}$ measurements include differences between the data and the MC simulation for the tracking efficiency and PID,
which can be investigated and corrected using a high purity control sample of about 20 M $Z\gamma$ (ISR return) events.

In the measurement of cross sections, the important uncertainties are from ISR correction factor
($1+\delta$, defined as the ratio of observed cross section over the Born one), luminosity measurement,
the branching fraction of intermediate state decay ($Z\rightarrow\mu^{+}\mu^{+}$), as well as fitting procedure.

The uncertainty of $1+\delta$ depends on the precisions
of both the experimental line shape measurement on the Born $\sigma_{ZH}$ below 250 GeV and the theoretical radiator function.
We expect the latter will be calculated to a precision that is negligible comparing to the statistical uncertainty by the time of the CEPC data taking.
The former is estimated with Born cross sections at six center-of-mass energies equally distributed from the threshold to 250 GeV.
The luminosity is set at 50 $fb^{-1}$ for each energy point
below 250 GeV and the cross sections are generated according to the formula in Ref.~\cite{samples} with statistical uncertainties.
The generated cross sections are fitted with the same formula,
but the coupling is free and the Higgs mass is float in one standard deviation according to PDG~\cite{pdg}.
Then the resultant line shape is used to calculate the $1+\delta$.
We repeat the generation, fitting, and calculation procedure 1000 times,
and the spread of $1+\delta$ is determined to be 0.1\% and taken as the systematic uncertainty of correction factor.

The integrated luminosity could be measured using small angle radiative Bhabha scattering and the expected precision is better than $10^{-3}$.
The current uncertainty in the ${\cal B}(Z\rightarrow \mu^{+}\mu^{-})$ is 0.2\%~\cite{pdg}, which will be further improved by the Z boson samples at the CEPC.
The uncertainty of fitting procedure could be estimated by changing the background shape and fitting range,
and the difference in the measured $\sigma_{ZH}$ is taken as the systematic uncertainty.

In the Higgs mass measurement, the dominant uncertainty may be from beam energy measurement. In order
to control this uncertainty to MeV level, the new technology needs to be developed to improve the precision of beam energy measurement.
Another potential uncertainty is the mass shift between the measured Higgs mass and the truth value.
In order to control the shift, the dependence of mass shift on the Higgs mass input is investigated around 125 GeV. Then it is
extracted as a third order Chebychev polynomial function. The measured Higgs mass is corrected by the fit function.
The combined uncertainty of fit function and the remaining shift, 1.5 MeV, is taken as the systematic uncertainty.

The consistence between the fast and full simulation is checked using the $ZZ$ events with at least one pair of muons found. The invariant mass of $\mu^+\mu^-$ is shown in Fig.~\ref{fastsimu}. In the concerned region between 80 GeV and 100 GeV , the statistics of the full-simulated ZZ sample is 2.32\% lower than that of the fast-simulated. If the
remaining background is reduced by 2\% after the final event selection, the precision of $\sigma_{ZH}$ is varied from 0.974\% to 0.971\% and the $m_{H}$ precision varied
from 6.91 MeV to 6.87 MeV. Therefore this effect can be safely ignored in this analysis.

\begin{center}
\centering
\includegraphics[width=7.5cm]{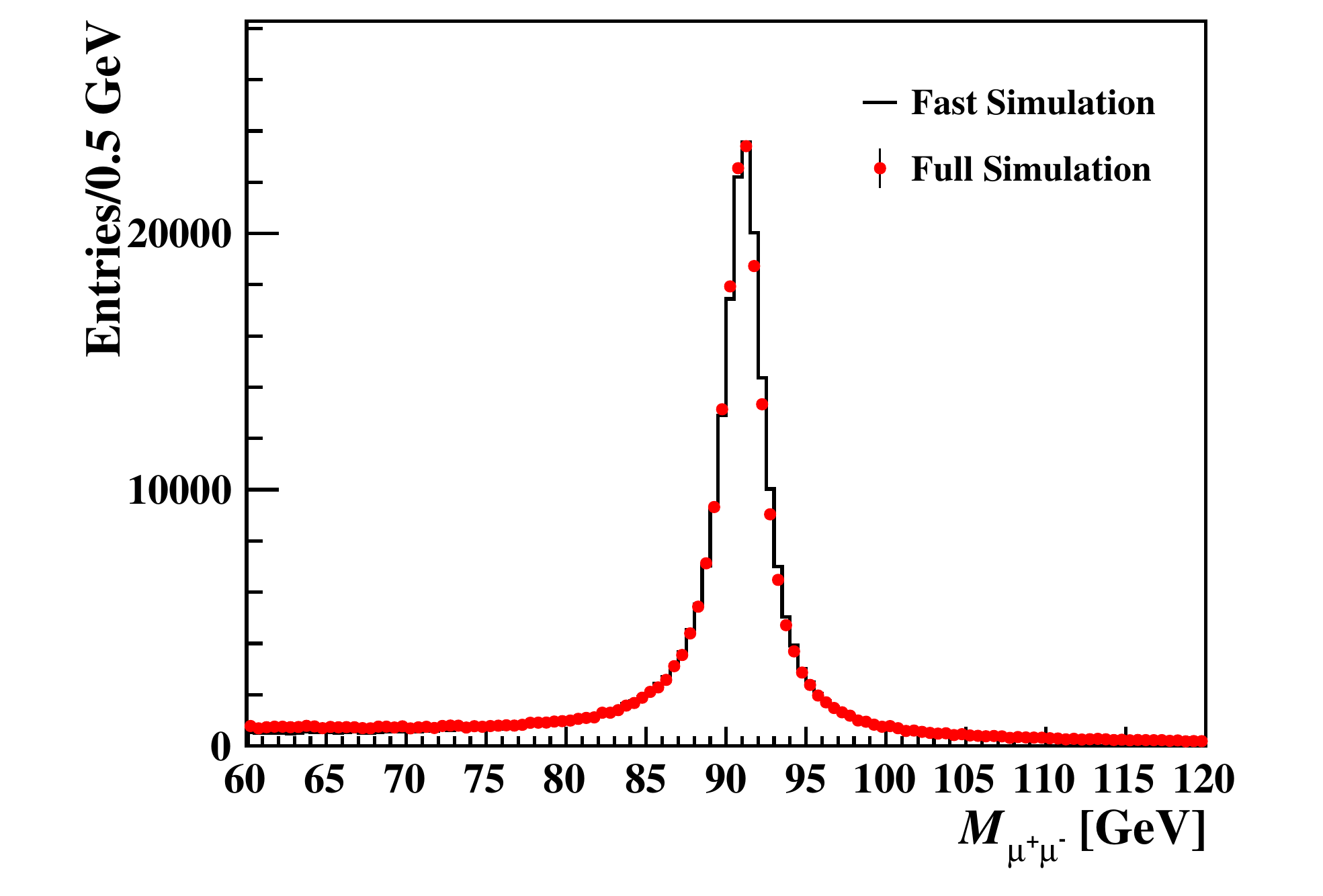}
\figcaption{The invariant mass spectrum of $\mu^{+}\mu^{-}$ from the
  samples of ZZ background. The red dots with error bars are full simulated while the black histogram is from fast simulation.}
\label{fastsimu}
\end{center}

From the above discussions, the systematic uncertainty should be under control while the statistical one will be dominated at the CEPC.

%\begin{multicols}{2}

\section{Summary}

The CEPC is expected to play a crucial role in understanding the nature of the Higgs boson.
In this paper, the statistical precisions of Higgs production cross section $\sigma_{ZH}$ and mass $m_{H}$ measurements at the CEPC are investigated
with full simulated Higgsstrahlung signal of 5 ab$^{\rm{-1}}$ integrated luminosity at the center-of-mass energy of 250 GeV.
Using the recoil mass method, the statistical precision of $\sigma_{ZH}$ could reach 0.97\%, corresponding to a 0.49\% accuracy of $g_{HZZ}$.
The expected statistical accuracy of $m_{H}$ is 6.9 MeV while it is improved to 5.4 MeV with inclusion of the Higgs decay information.
The dependence of these results on TPC radius is investigated and parameterized.
Reducing the TPC radius by 25\%, the statistical precisions of $\sigma_{ZH}$ and $m_{H}$ are worsened to 0.98\% and 8.4 MeV, respectively.
In addition, we explored the potential of the invisible decay mode of the Higgs boson at the CEPC.
The upper limit of ${\cal B}(H\rightarrow \rm{inv.})$ at the 95\% confidence level could reach 1.2 $\times$ 10$^{\rm{-2}}$.
All above results are incorporated into the CEPC-SPPC Preliminary Conceptual Design Report~\cite{CEPC}.

The same measurement is studied at the ILC in the $Z\rightarrow e^+e^-$ and $Z\rightarrow \mu^{+}\mu^{-}$ channels with an
integrated luminosity of 2 ab$^{\rm{-1}}$ and polarized beams of $P(e^{-},e^{+})=(-0.8,0.3)$\cite{ILC,ilcrecoil2}. The $g_{HZZ}$ precision could reach 0.4\% while
the upper limit of ${\cal B}(H\rightarrow \rm{inv.})$ is 1.7 $\times$ 10$^{\rm{-2}}$. For the $m_{H}$ measurement,
the current $m_{H}$ precision is 0.24 GeV achieved at the LHC~\cite{lhcsub5} and it will be improved to 50 MeV at the HL-LHC~\cite{hllhc}.
At the ILC, the statistical precision of $m_{H}$ could reach 14 MeV~\cite{ILC,ilcrecoil2}.
Compared with these facilities, the improvement at the CEPC is due to weaker beamstrahlung and higher statistics.

\acknowledgments{The authors would like to thank the ILD Concept Group for providing a reference of detector and software for the CEPC study. We thank professor
Yuan-Ning Gao for fruitful discussion on the analysis technique.
We are grateful to Dr. Xin Mo and Yu-Qian Wei in providing high statistical MC samples.
We appreciate Dr. Bin-Song Ma on the development of reconstruction algorithm.

%This work is supported by the Joint Fund
%s of the NSFC under Contracts No. U1232105
%and funding from CAS for the Hundred Talent programs No. Y3515540U1.
}

\end{multicols}

%\vspace{10mm}

\vspace{2mm}
\centerline{\rule{80mm}{0.1pt}}
\vspace{2mm}

\begin{multicols}{2}

\end{multicols}

\clearpage

\end{CJK*}
\end{document}